# Physical Study by Surface Characterizations of Sarin Sensor on the Basis of Chemically Functionalized Silicon Nanoribbon Field Effect Transistor

K. Smaali[1,§], D. Guérin[1], V. Passi[1], L. Ordronneau[2], A. Carella[2], T. Mélin[1], E. Dubois[1], D. Vuillaume[1], J.P. Simonato[2] and S. Lenfant[1,*]

[1] IEMN, CNRS, Avenue Poincaré, Villeneuve d'Ascq, F-59652 cedex, France.

[2.] CEA, LITEN/DTNM/SEN/LSIN, Univ. Grenoble Alpes, MINATEC Campus, F-38054 Grenoble, France.

**ABSTRACT :** Surface characterizations of an organophosphorus (OP) gas detector based on chemically functionalized silicon nanoribbon field-effect transistor (SiNR-FET) were performed by Kelvin Probe Force Microscopy (KPFM) and ToF-SIMS, and correlated with changes in the current-voltage characteristics of the devices. KPFM measurements on FETs allow (i) to investigate the contact potential difference (CPD) distribution of the polarized device as function of the gate voltage and the exposure to OP traces and; (ii) to analyze the CPD hysteresis associated to the presence of mobile ions on the surface. The CPD measured by KPFM on the silicon nanoribbon was corrected due to side capacitance effects in order to determine the real quantitative surface potential. Comparison with macroscopic Kelvin probe (KP) experiments on larger surfaces was carried out. These two approaches were quantitatively consistent. An important increase of the CPD values (between + 399 mV and + 302 mV) was observed after the OP sensor grafting, corresponding to a decrease of the work function, and a weaker variation after exposure to OP (between - 14 mV and - 61 mV) was measured. Molecular imaging by ToF-SIMS revealed OP presence after SiNR-FET exposure. The OP molecules were essentially localized on the Si-NR confirming effectiveness and selectivity of the OP sensor. A prototype was exposed to Sarin vapors and succeeded in the detection of low vapor concentrations (40 ppm).

# 1. INTRODUCTION

The functionalization of semiconductor surfaces with organic molecules has become an increasingly popular topic because of its potential in many applications, especially gas sensors. To improve the sensitivity of these sensors based on functionalized semiconductors, a preferred approach is to increase the surface to volume ratio by using nanostructures like nanowires[1;2;3] or nano-ribbons. This allows to maximize the amount of molecules or biomolecules immobilized by volume for a given planar surface of semiconductor[4]. Kelvin Probe Force Microscopy (KPFM) is of a great interest to understand quantitatively the electronic properties of these nano-structures with a high precision[5]. KPFM is capable to map the contact potential difference (CPD) at the nanometer scale, it gives access to quantitative measurements of the electronic properties of a functionalized surface by Self Assembled Monolayers (SAM)[6;7;8], DNA[9] or organic nanostructures[10;11]. These electronic properties include the work function[12], i.e. the energy gap between the Fermi level and the vacuum level, and dipole moments[13]. KPFM has also been used frequently on single nanowire to characterize Ohmic and Schottky contacts[14], localized trapped charges[15;16;17;18;19], dopant distributions[20;21] or Schottky junction depletion region[22]. KPFM presents several advantages such as (i) a quantitative measurements of the CPD value at the nanometer scale with a potential sensitivity around few mV; (ii) it is a nondestructive technique; (iii) direct characterization of polarized device without important perturbation due to the absence of contact between the tip and the surface[23]; (iv) applicability on conducting and insulating substrates.

Recently, we have demonstrated a highly sensitive (sub ppm) and selective nerve agent sensor based on electrical transduction of a chemical reaction occurring on the surface of a functionalized Silicon NanoRibbon Field Effect Transistor (SiNR-FET)[24;25]. Better sensitivity was obtained on these sensors by reducing wire width from a few µm to 25 nm[26;27]. Thanks to these size modifications which increased the surface to volume ratio of the conductive channel, the current gain of the SiNR-FET was improved by 4 decades. In the present paper, we report on a direct measurement of the CPD at the nanometer scale by performing KPFM study on SiNR-FET biased at the operating voltages. These results are compared with Kelvin Probe measurements at the macroscopic scale. The SiNR surface was modified by a SAM selectively sensitive to organophosphorus gases. The diphenyl chlorophosphate DPCP was used as a simulant of OP nerve agents. The drain-source current ($I_{DS}$) plotted versus the gate voltage ($V_G$) shows the ambipolar behavior of the FET. The exposure of functionalized SiNR-FET to vapors of DPCP induced a strong modification of the $I_{DS}$- $V_G$ curve[24]. In the following, a more detailed characterization of the SiNR-FET is presented in order to explain the observed electrical properties. The main aims of this work are to associate different surface characterization techniques as KPFM, Kelvin Probe and ToF-SIMS and electrical measurements in order to (i) understand the physical and chemical effects occurring on the sensor surface; (ii) optimizing the sensor and (iii) demonstrate the potential of the KPFM technique for electrical device studies. Our study focuses on three aspects: (i) an electrostatic aspect for the different chemical steps ("naked" SiNR, after surface functionalization by the SAM, and after DPCP exposure) by comparing the CPD mapping measured by KPFM on the biased SiNR-FET with Kelvin Probe measurement realized on large surfaces with the same chemical steps; (ii) a chemical speciation using ToF-SIMS 2D element mapping experiments to localize precisely the reaction of molecules on the surface of the Si-NR FET for each chemical step; (iii) an applicative aspect with the demonstration of a prototype evaluated in real conditions with sarin, a wellknown chemical warfare agent.

## 2. MATERIALS AND METHODS

**Device fabrication.** Silicon Nano-Ribbons (Si-NR) were formed from silicon on insulator (SoI) wafers using e-beam lithography and dry reactive-ion etching steps, following the procedure described elsewhere[24] (see also supporting information). The structure of the SiNR-FET (figure 1a) is based on a Si-NR (70 nm thick, (100)-oriented, p-doped with Boron at $10^{15}$ atom.cm$^{-3}$) with different lengths and widths (4 x 4 μm; 4 x 1 μm) connected by source and drain Ti/Au (10/100nm) contacts patterned by e-beam lithography and a lift-off process. The thickness of the buried oxide (BOX) of the SOI wafer is 140 nm, and serves as the gate dielectric layer combined to a back-gate command through the silicon wafer handler. The SiNRs were functionalized by covalent grafting through thermal hydrosilylation of **1** onto HF-pretreated substrate (more details in the supporting information). Compound **1** named 3-(4-ethynylbenzyl)-1,5,7-trimethyl-3-Azabicyclo[3.3.1]nonane-7-methanol is hereafter referred to as TABINOL. Diphenylchlorophosphate (or DPCP) was used as a simulant of nerve agents (sarin) because of its similar structure and chemical reactivity, but much lower toxicity. TABINOL reacts with DPCP and forms azaadamantane quaternary ammonium salt following the reaction presented in figure 1b (more information for the chemistry synthesis and reaction are detailed elsewhere[24]). For a more detailed surface characterization, large pieces (1cm x 1cm) of silicon wafers were also functionalized with TABINOL and exposed to DPCP using the same protocols.

**Contact angle measurements.** We measured the water contact angle (on the 1 cm² sample) with a remote-computer controlled goniometer system (DIGIDROP by GBX, France). We deposited a drop (10-30 µL) of deionized water (18 MΩ.cm) on the surface and the projected image was acquired and stored by the computer. Contact angles were extracted by a contrast contour image analysis software. These angles were determined a few seconds after application of the drop. These measurements were carried out in a clean room (ISO 6) where the relative humidity (50%) and temperature (22°C) are controlled. The precision of these measurements is ± 2°.

**Thickness measurements.** We recorded spectroscopic ellipsometry data (on the 1 cm² sample) in the visible range using an UVISEL (Horiba Jobin Yvon) Spectroscopic Ellipsometer equipped with DeltaPsi 2 data analysis software. The system acquired a spectrum ranging from 2 to 4.5 eV (corresponding to 300 to 750 nm) with intervals of 0.1 eV (or 15 nm). Data were taken at an angle of incidence of 70°, and the compensator was set at 45.0°. We fitted the data by a regression analysis to a film-on-substrate model as described by their thickness and their complex refractive indexes. We estimated the accuracy of the SAM thickness measurements to ± 1 Å.

**ToF-SIMS.** ToF-SIMS spectra measurements and images were carried out in positive and negative modes using a ToF-SIMS V instrument (ION-TOF GmbH Germany) on the Si-NRFET. This instrument is equipped with a Bi liquid metal ion gun (LMIG). Pulsed $Bi_3^+$ primary ions have been used for analysis (25 keV) in both bunch and burst alignment modes. Spectra and images were taken from an area of 35µm x 35µm. In bunch mode, mass resolution was > 8500 at m/z= 196, 249, 294, 312 and 314, for $C_{12}H_{22}NO^+$, $C_{12}H_{10}PO_4^-$, $C_{21}H_{28}N^+$, $C_{21}H_{30}NO^+$, $C_{21}H_{32}NO^+$, respectively.

**KP measurements.** Contact potential difference (CPD) was measured on large pieces (1 cm x 1 cm) of wafers/samples using a macroscopic Kelvin probe (KP) system (from Kelvin Probe Technologies) in ultra-high vacuum ∼ 4-5x10$^{-10}$ Torr at 24°C. The Kelvin probe technique measures the contact potential difference between two surfaces brought in close proximity. For sake of clarity and consistence with respect to Kelvin probe force microscopy experiments described hereinafter, we define the CPD as:

$$CPD = (W_{tip} - W_s)/e \qquad\qquad\qquad\qquad\qquad\qquad\qquad\qquad (Eq.\ 1)$$

with $W_s$ and $W_{tip}$ are the work functions of the sample and the tip respectively, and e the absolute magnitude of electron charge (see SI for experimental details). In the actual Kelvin probe instrument a metallic tip is vibrated at a frequency of 80 Hz. Using a high-gain, low noise amplifier, the AC(ω) current generated by the oscillation is monitored. Gradual ramping of the counter potential and finding zero AC current gives CPD.

**KPFM measurements.** On the Si-NR FET, we measured, locally, CPD on the silicon channel by Kelvin Probe Force Microscopy (KPFM). KPFM measurements were carried out at room temperature with a Dimension 3100 from Veeco Inc., purged with a flow of dry nitrogen atmosphere and controlled by the Nanoscope v5.30R2 software. Images were processed using WSxM 5.0 software from Nanotec Electrónica[28]. We used Pt/Ir (0.95/0.05) metal-plated cantilevers with spring constant of ca. 3 N/m and a resonance frequency of ca. ω/2π = 70 kHz. Topography and KPFM data were recorded using a standard two-pass procedure[29], in which each topography line acquired in tapping mode is followed by the acquisition of KPFM data in a lift mode, with the tip scanned at a distance z ∼ 80 nm above the sample so as to discard short range surface forces and be only sensitive to electrostatic forces. DC and AC biases ($V_{DC} + V_{AC} \sin(\omega t)$) are applied to the cantilever with $V_{AC}$ = 2 V. Experimentally, the contact potential difference (CPD) is measured using a feedback loop which sets to zero the cantilever oscillation amplitude at ω by adjusting the tip DC bias $V_{DC}$. Note that for KPFM measurements on a voltage biased Si-NR FET, this value is an effective CPD which has to be corrected from the side capacitance effects[30]. This corrected CPD is noted $CPD^*$ thereafter. In our case, given the AFM tip and the Si-NR FET geometry the $CPD^*$ value corrected by side capacitance effects at the Si-NR center can be estimated as (see SI):

$$CPD^* = 1.75 (CPD - 0.27\ V_{ox} - 0.16\ V_{Au}) \tag{Eq. 2}$$

with $V_{ox}$ the CPD of the oxide and $V_{Au}$ the CPD on the Au electrodes. $V_{ox}$ and $V_{Au}$ are measured on the same device by putting the KPFM tip on the oxide beside the Si NR and source and drain electrodes, respectively.

The work function (WF) of the surface ($W_s$) is deduced from the $CPD^*$ following the relation:

$$W_s = W_{tip} - e\ .\ CPD^* \tag{Eq. 3}$$

where *e* is the elementary charge and $W_{tip}$ the work function of the KPFM tip (for Pt/Ir tip $W_{tip}$ = 4.28 ± 0.07 eV[31]). We can notice here than the same definition was adopted between the CPD measured by KP and KPFM (Eq. 1 and Eq. 3).

During the KPFM measurements, SiNR-FETs were electrically characterized in situ using an Agilent 4155C Semiconductor Parameter Analyzer. Electrical contacts on the source and drain electrodes of the SiNR-FET were realized using micro-probers connected to the Agilent 4155C. The silicon back gate of the Si-NR FET was glued with a conductive silver paste on a metallic handler electrically connected to the AFM chuck. The chuck was connected to the Agilent 4155C. The drain current - gate voltage characteristics, $I_{DS}$-$V_G$, were obtained by sweeping the gate voltage ($V_G$) and keeping the drain-source voltage ($V_{DS}$) at a constant value. This protocol allowed us to obtain current-voltage curves and local CPD mapping simultaneously. To analyze the effect of $V_G$ on the CPD of the Si-NR at a fixed $V_{DS}$, the same segment of the Si-NR with both electrodes (110 µm length) (corresponding to a line in the KPFM image) was measured by KPFM as a function of time (the time corresponding of the ordinate of the KPFM image: function "slow scan axis" disabled in Nanoscope Software). Images were acquired by sweeping the gate voltage from -10 V to 10 V and then from 10 V to -10 V with a 1 V step at a fixed $V_{DS}$. Each value of $V_G$ corresponds approximately to 10 lines in the KPFM image. Values on Si-NR presented at a fixed $V_G$ were an average of about 23 pixels corresponding to the length of the Si-NR, and an average of these pixels on the 10 lines obtained at the same $V_G$.

We also recorded I-V and KPFM sequentially. In this case, I-V characterizations were first recorded in a glove box filled with clean nitrogen ($O_2$ < 1 ppm, $H_2O$ < 1 ppm) and then transferred to the Dimension 3100 for KPFM characterization with no applied bias on the SiNR-FET. These measurements were done on the same device before and after each chemical steps, i.e., we measured: i) the "naked" Si-NR after HF passivation, ii) the Si-NR functionalized with TABINOL before DPCP exposure; and iii) after DPCP exposure.

## 3. SURFACE CHARACTERIZATIONS

### 3.1 Water contact angle and thickness

Water contact angle and thickness measurements are quick methods to assess the effectiveness of the chemical reactions on the Si surface. Table 1 gives the results obtained after TABINOL functionalization and exposure to DPCP.

The thickness of 1.65 nm corresponds roughly to a monolayer of TABINOL since the length of this molecule is about 1.4 nm (from geometry optimization calculation with MOPAC - see crystallographic data[24]). Since the naked Si-H surface (after cleaning) is highly hydrophobic (water contact angle ~90°[32]), the decrease of the water contact angle to 66° is consistent with the presence of a more hydrophilic group at the outer surface, mainly the primary alcohol group. Upon reaction with DPCP, the thickness is slightly increased (depending on the duration of DPCP exposure) and the water contact angle decreases slightly. These results are consistent with previous experiments.[24]

### 3.2 KP and KPFM measurements on large samples

Since the top surface of the nanoribbons is (100) and sidewalls are (110) orientated, and to compare with localized KPFM on the Si-NR FETs (see next section), we measured the CPD on large samples for these two orientations, and also on (111) surfaces for the sake of completeness. Figure 3 shows the results for the naked Si surface (just after oxide removal and cleaning), after grafting TABINOL and after exposure to room temperature DPCP vapor pressure (for 1 h). Kelvin probe measurements can only provide relative values, i.e. difference of work function between the tip and surface under study. This variation of work function is related to the CPD defined in Eq. 1. We defined the relative values, noted $\Phi$ (i.e. $\Phi_{TAB}=CPD_{TAB} - CPD_{ref}$ and $\Phi_{DPCP}= CPD_{DPCP}-CPD_{TAB}$), with the reference samples being the naked Si surface.

Firstly, we note that there is a difference in the CPD values when similarly treated samples are compared for different orientations. This variation is attributed to the fact that the surface termination of silicon for the different orientations is not the same, i.e., formation of di-hydrides, mono-hydrides and a combination of mono- and tri-hydrides on (100), (110) and (111) orientations, respectively.[33]. For the naked Si surfaces, these different chemical environments lead to variations in the Si work function[34; 35]. After grafting the TABINOL, the CPD values are increased, again with relative amplitude which is orientation dependent. Grafting molecules on Si surfaces is well known to induce interface dipoles which modulate the work-function.[36;37;38] This work-function variation depends on the density of molecules and on the surface orientation, and it is also known that the density of grafted molecules depends on the Si surface orientation.[39;40;41;42] However, for the two Si orientations of interest in this work, i.e. (100) and (110), the increase of the CPD after the grafting of TABINOL is almost the same (Figure 3). After exposure to DPCP, the variations of CPD are weak.

In order to compare the KP and KPFM measurements, large surfaces (1 cm x 1 cm) of silicon substrate were measured before and after TABINOL grafting by KP and KPFM (Table 2). The grafting of TABINOL in both cases, induces an increase of the CPD of $\Phi_{TAB} = CPD_{TAB} – CPD_{REF}$ ~ + 350 mV and ~ + 399 mV for KPFM and KP respectively, with $CPD_{REF}$ and $CPD_{TAB}$ the measured CPD on naked Si-NR and after TABINOL grafting, respectively. This variation of 49 mV (14% of the CPD value) between the KP and KPFM measurements can be explained by a variation of the silicon oxidation of the Si surface (see in SI part 6 the air exposure effect on the CPD measured by KP). In KPFM (or KP), the positive (or negative) values for the CPD correspond to a lower (or higher) WF for the sample than the tip. In both techniques, the variation measured after TABINOL grafting, was associated to an increase of the CPD, corresponding to a decrease of the sample work function. These variations measured by the two instruments are in good agreement. These results validate the comparison between measurements by KP and KPFM, which are mutually consistent.

**3.3 ToF-SIMS**

Since the surface characterization techniques (contact angle, ellipsometry) are not applicable to the Si-NRFET (too small surfaces), we used 2D mapping ToF-SIMS experiments to check the surface functionalization of the nanowire transistors (after TABINOL grafting) and to assess the reaction of tabinol with DPCP. Thanks to molecular imaging mass spectrometry (see Fig. 4), we could highlight precisely the localized grafting of TABINOL on the silicon nanoribbon and then the reaction with DPCP by the observation of certain fragment ions resulting from TABINOL ($C_{12}H_{22}N^+$, $C_{12}H_{22}NO^+$, $C_{21}H_{32}NO^+$) and from fragmentation of the organophosphorus compounds ($PO_3^-$, $PO_2^-$). Indeed, as specified in Table S1 (supporting information), characteristic TABINOL fragment ions were observed before and after DPCP whereas $PO_3^-$ and $PO_2^-$ ions were only observed after exposure to organophosphorus molecules. Moreover, molecular imaging (Fig. 4) shows that these latter fragment ions are selectively localized on the Si-NR (4×4 µm) confirming the effectiveness of the DPCP reaction on NR-Si FET functionalized with TABINOL. Note that some of these fragments were also observed on the neighboring gold contact electrodes since amine and phosphorus derivatives are also known to readily chemisorb on this metal. However, the most intense signal comes clearly from the Si-NR region (yellow spots in Fig. 4).

## 4. KPFM AND I-V RESULTS ON SI-NR FET AND DISCUSSION

**4.1 Sequential approach**

In a first step, I-V and KPFM measurements were done sequentially as described above. These measurements were done to establish the first trends of the effects of the chemical treatments on the electrical behaviors of the devices. We present in the following the IV and KPFM results obtained on Si-NR with the larger surfaces (4 x 4 µm²).

Transfer curves of the ambipolar SiNR-FET are highly dependent of the surface functionalization of the Si-NR (figure 5). After TABINOL grafting, the main effect is a shift towards more negative voltages. Here the voltage shift is calculated from the voltage position, $V_m$, of the minimum current in the IV curves. We have $\Delta V_m \sim$ 4.2 V and 4.4 V for the forward and reverse bias sweeps, respectively. The exposition to DPCP vapor induced: (i) a shift of the curves to positive gate voltages (shift of $\Delta V_m$ = 3.3 V and $\Delta V_m$ = 4.4 V, respectively, for the forward and reverse bias sweep as shown in figure 5); (ii) an increase of the hysteresis behavior. The shift of the IV curves may be due to several factors: (i) charge trapping/creation in the SAMs or at the SAM/Si-NR interface; (ii) charge transfer between silicon and molecules (induced interface dipole), (iii) modification of molecular dipole. Each of these three factors leads normally to a change in the silicon SP of the Si-NR, and KPFM is used to gain more insights in the physical origins of the IV shift.

KPFM measurements, as expected, show significant changes of the CPD of the Si-NR with the chemical functionalization (figure 6). KPFM images and CPD profiles show a clear homogeneity on the Si-NRs. The experimental CPD (obtained by averaging on the Si-NR in Fig. 6) were corrected using Eq. 2, with values of $V_{ox}$ and $V_{AU}$ extracted from the KPFM image (Fig. 6) on the oxide region and on the gold electrode, respectively. We can notice than $V_{AU}$ remains stable after TABINOL grafting ($V_{AU} \sim$ +36 mV and $V_{AU} \sim$ +40 mV before and after TABINOL grafting respectively) and weak shifts after DPCP exposure ($V_{AU} \sim$ +85 mV after DPCP exposure). This CPD shift with the DPCP exposure is explained by the chemisorption of few DPCP molecules on the gold electrode as observed by ToF-SIMS (see section 3.3). With the CPD correction (using Eq. 2), a significant increase $\Phi_{TAB}$ = CPD*$_{TAB}$ - CPD*$_{REF}$ ~ + 302 mV of the CPD* with the TABINOL grafting is observed, and a weak decrease of $\Phi_{DPCP}$ = CPD*$_{DPCP}$ -

CPD$^*_{TAB}$ ~ - 61 mV is measured after exposure to DPCP, with CPD$^*_{REF}$, CPD$^*_{TAB}$ and CPD$^*_{DPCP}$ being the corrected CPD (with Eq. 2) on naked Si-NR, after TABINOL grafting, and after DPCP exposure, respectively. The same study was systematically repeated for various Si-NR FETs with other Si-NR geometries. On these various devices, dispersion was observed on CPD values (variations of 80% in maximum) and on IV characteristics. This dispersion may be explained by the uncertainty on the oxidation of silicon; KPFM measurements are done in $N_2$ purged atmosphere, thus likely with residual oxygen. But in most cases, the same trends on the CPD than those presented before were observed.

If we compare these measurements on Si-NRFET with those on large sample by KPFM (section 3.2), we obtained $\Phi_{TAB}$ ~ + 302 mV $\Phi_{TAB}$ ~ + 350 mV for the Si-NR FET and for large surface, respectively. These CPD drops with TABINOL grafting are comparable: (i) same sign: the grafting of TABINOL is associated to an increase of the CPD, corresponding to a decrease of the work function (Eq. 3); (ii) a weak variation (around 14%) on $\Phi_{TAB}$ could be explained by the fact than the density of grafted TABINOL molecules is not exactly the same for large surface and Si-NR. This good consistency between measurements on microstructures and on large surfaces validates the side capacitance correction (section SI-4) with the use of Eq. 2.

To establish a correlation between the IV curves and the KPFM experiments, the effect of the chemical steps (TABINOL grafting and then DPCP exposure) was determined using a couple of parameters: $\Delta V_m$ (shift of the current minimum picked from the $I_{DS}$-$V_G$ curve) and $\Phi$ (change in the CPD). Since we have measured the back and forth IV curves, we considered two voltage shifts: $\Delta V_{mfor}$ and $\Delta V_{mrev}$ for the forward and reverse IV traces, respectively. The grafting of a TABINOL monolayer induces a drop of the CPD$^*$ in the $\Phi_{TAB}$ ~ + 302 mV range and a voltage shift of the IV curves in the 4.2-4.4 V range (Fig. 5). Note that the IV hysteresis (in Fig. 5 the difference between $\Delta V_{mfor}$ and $\Delta V_{mrev}$) remains the same as for the "naked" Si-NR FETs, i.e. of the order of 3.6-3.8 V. On the contrary, DPCP expose induces a weak CPD$^*$ change, $\Phi_{DPCP}$ = -61 mV, but a significant larger IV hysteresis (ca. 4.7 V) (Fig. 5). Since the TABINOL grafting does not increase the IV hysteresis, we can discard the trapping of charges as the origin of the electrical modifications of the Si-NR FETs. Thus, we assume that the decrease of CPD is mainly due to the dipole moment of the TABINOL grafted on the silicon surface. We can estimate this dipole using the Helmholtz equation:

$$\Phi_{TAB} = \frac{NP_z}{\varepsilon_0 \varepsilon_{SAM}}$$ (Eq. 4)

where N is the surface density of molecules in the SAM, estimated around $2.5 \times 10^{14}$ cm$^{-2}$ from previous work[43], $P_z$ is the dipole moment of the molecule perpendicular to the surface, $\varepsilon_0$ is the vacuum dielectric permittivity, $\varepsilon_{SAM}$ is the relative permittivity of the SAM (here 2.5). From Eq. 4 we estimate a value of dipole moment perpendicular to the surface $P_z \sim 0.81$ D. Since, the KPFM measurement shows a decrease of the WF after TABINOL grafting, this dipole has its positive side pointing out the TABINOL SAM. This is consistent with the presence of tertiary amines at the end of the TABINOL. The dipole moment for a single TABINOL molecule in vacuum is estimated $\sim 0.44$ D with the MOPAC software (ChemOffice Software). Considering the classical electrostatics of the FET device, a potential change of $\Phi$ at a distance d (through a material with a dielectric constant $\varepsilon_{SAM}$) from the Si channel (virtual top gate) is roughly equivalent to a change $\Delta V_G$ of the back- gate voltage through a gate oxide of thickness $t_{ox}$ (dielectric constant $\varepsilon_{ox}$) according to $\Delta V_G = (t_{ox}/d)(\varepsilon_{SAM}/\varepsilon_{ox}) \Phi$. With $t_{ox}$ = 140 nm, $e_{ox}$ = 3.9, d= 1.7 nm (from thickness measured by ellipsometer, see section 3.1) a back-gate shift of 4.2-4.4 V roughly corresponds to $\Phi$ of 80-84 mV, i.e. of the same order of magnitude as the KPFM measurements.

After DPCP exposure, the situation is different. We have measured a weak CPD change from KPFM, $\Delta\Phi$ = - 61 mV, but a sensitively larger IV hysteresis (ca. 4.7 V). During DPCP exposure, TABINOL reacts with DPCP and produces azaadamantane quaternary ammonium salt and diphenyl phosphate counter ions on the Si-NR FET surface[24]. Positive charges (N$^+$) created on top of the monolayer after treatment with DPCP (see Fig. 1) can eventually modify the CPD. However, a negatively charged counter-ion is also created and the monolayer is likely to remain neutral with no further contribution to the CPD. Diphenyl phosphate counter ions are randomly oriented on the surface, and, in average, we no dot expect a contribution of its specific dipole to the CPD. These features are consistent with the weak $\Phi_{DPC}$ measured by KPFM, as well as by KP and KPFM on large surfaces. "Mobile" phosphonate ions present in the SAM after DPCP exposure can cause the more important hysteresis observed in the IV characteristics (4.7 V in Figs. 5).

**4.2 Direct measurement by KPFM of polarized Si-NR FET**

*4.2.1 Effect of $V_G$ on the SP of Si-NR at $V_{DS}$ = 0 V*

To analyze more precisely these charge effects, KPFM were performed on the Si-NR FETs biased by different back gate ($V_G$) and source-drain ($V_{DS}$) potentials. In the case of "naked" Si-NR before any chemical step, the effect of the back gate ($V_G$) on the SP measured by KPFM was studied for a source-drain potential $V_{DS}$ = 0 V (figure 7). For $V_G$ = 0 V, the CPD profile is similar to the previous one presented in figure 6 with a CPD difference between the SI-NR "naked" and gold comprised between 360 - 390 mV. At $V_G \neq 0$, the smooth transition region in the CPD profile outside the Si-NR region extending over the gold electrodes corresponds to side capacitances between the device surface and the KPFM tip[29] (see SI). The length of this region (around 50 µm) is very close to the total width of the AFM tip comprised between 45 – 55 µm according to technical data, showing that the capacitance formed by the different surfaces of the device and the sides of KPFM tip is not negligible. KPFM signals have to be corrected to estimate the real CPD (Eq. 1).[44;45] Fig. 7-b shows a linear variation of the uncorrected effective CPD (value measured on the middle of Si-NR channel) with $V_G$. By using Eq. 1 on the CPD values in Fig. 7-b, the corrected CPD* obtained is not significant, in view of the low values of CPD measured in absence of $V_{DS}$. In that later case, we do not have any $V_G$ dependence of the CPD. It means that surface potential of the Si-NR is not modulated by the applied back-gate. A possible explanation would be that (i) the KPFM characterization system is not sufficiently sensible to detect variation of the CPD with the back gate potential ($V_G$) on Si-NR, or (ii) that the real effect is masked by the side capacitance effects. At the silicon/gold interface, we observe a clear voltage drop that reveals a Schottky contact formed by the gold electrode with the Si-NR[14]. Moreover, peaks are present mainly at the right interface, and the peak height increase with the bias (peak height up to 350mV for $V_G$ = 10V). A possible origin of these peaks is due to the presence of localized charges at the Si-NR/electrode interface, caused by a mismatch between metallic layers and consequently by the presence of oxidized Ti locally.

A KPFM scan on a same segment of the "naked" Si-NR (inset in figure 8) versus $V_G$ and at $V_{DS}$ = 0 V (*i.e.* no charge transport through the Si-NR) shows the reversible effect of the back gate potential on the Si-NR effective CPD (figure 8). Starting on top of the image at $V_G$ = -10 V, the back gate voltage increases to $V_G$ = 10 V in the middle of the image, and finally decreases to $V_G$ = -10 V at the bottom of the image (step of 1V every 10 lines approximately). The two CPD profiles at $V_G$ = -10V (lines at the top and bottom of the image) are similar showing the reversible effect of $V_G$. We can notice here the absence of hysteresis between the forward bias (from -10V to 10V) compared to the reverse bias (from +10V to -10V) (the two half images are symmetric).

*4.2.2 Effect of $V_G$ on the SP of Si-NR at $V_{DS}$ = -2 V*

KPFM characterizations were subsequently performed by applying a source-drain potential ($V_{DS}$ = -2 V) on a "naked" Si-NR (figure 9-b). The main differences on CPD profiles between $V_{DS}$ = 0 V and $V_{DS}$ = -2 V are: (i) the shift of the right electrode CPD from 0 V (figure 8) to -2 V (figure 9-b) due to the polarization of this electrode; to measure exactly this difference of -2 V on CPD between source and drain electrodes, it is necessary to take the value measured at a distance of X = 0 µm and X = 110 µm of the Si-NR, due to the presence of the transition region (side capacitance effect as discussed above); (ii) the important variation of the CPD value for $V_{DS}$ = -2 V: for $V_G$ = +10 V (ON state of the Si-NR FET with an electron conduction), the maximum value in the middle of the NR is equal to $CPD_{Si-NR}$ = 5.97 V, for $V_{DS}$ = -2 V and for $V_G$ = -10 V (ON state of the Si-NR FET with a hole conduction), the minimum value is equal to $CPD_{Si-NR}$ = -6.86 V. Corrected from the side effects (Eq. 1), the corrected $CPD^*$ of the Si-NR varies between 5.7 V and -7.3 V (variations of the CPD value after corrections are inferior 6%). This important variation of KPFM signal observed here (compared to the absence of $V_G$-dependency at $V_{DS}$=0 (Fig. 7-b) is explained by the presence of carriers (electrons for positive $V_G$ and holes for negative $V_G$) into the Si-NR during the KPFM measurement.

Fig. 10 shows the CPD dependence of the Si-NR surface (average value on the Si-NR see methods) with $V_G$, and the corresponding IV data acquired simultaneously during the KPFM measurements (see methods). Here, the CPD* on Si-NR (CPD*$_{Si-NR}$) presents an important hysteresis. The maximum of the CPD* hysteresis was measured for $V_G$ = -3 V with a variation $\Delta CPD^*$ = 2,5V between the two curves. From the higher back gate potentials ($V_G$ = -10 V and 10 V), we can estimate that $V_G$ induces a variation on CPD*$_{Si-NR}$ comprised between 33-42 % of the applied $V_G$. So for $V_{DS}$ = -2 V, we clearly observe by KPFM measurement (i) a modification of the Si-NR CPD with the back gate potential ($V_G$) around 68 ± 7 % of $V_G$, and (ii) an important hysteresis of the CPD with $V_G$.

The transfer IV curve of the Si-NR FET (Fig 10) is qualitatively similar to that presented in figure 5 for the sequential method, with an ambipolar behavior: the Si-NR FET transistor is in ON state for negative $V_G$ < -4 V (conduction by holes) and for positive $V_G$ (conduction by electrons), and the Si-NR FET is in OFF state for $V_G$ ~ -2 to -3 V. In figure 10, the IV curves also display a small hysteresis, with its maximum of $\Delta V_G$ around 1V for gate voltages at the current minima ($V_m$ ≈ -3 and -2V). This maximum of the IV hysteresis is observed at approximately the same potential as for the maximum on the Si-NR CPD hysteresis (at $V_G$ = -3 V see above). This hysteresis behavior can be due to trapping and detrapping of charges at the Si-NR surface caused by the presence of traps at the molecule/Si interface.[46;47]

*4.2.3 Effect of the chemicals functionalization on SP*

The comparison of the Si-NR FET transfer curve and CPD dependence with $V_G$ acquired simultaneously on the Si-NR FET functionalized by TABINOL, before and after DPCP exposure shows clearly an effect of the simulant of nerve agents on the measured properties (Fig. 11). For the transfer curves, the same behavior as previously described (Fig. 5) is observed with the DPCP exposure: a shift of the curve to positive gate voltage bias (shift $\Delta V_m$ = 2 V and $\Delta V_m$ = 3 V for the forward and reverse bias respectively). For the CPD dependence with $V_G$, as observed on "naked" Si-NR at $V_{DS}$ = -2 V, the CPD on Si-NR is not perfectly linear with $V_G$, and presents a hysteretic behavior. DPCP increases the CPD hysteresis between the forward and reverse bias measured $\Delta CPD$ = 0.47 V and $\Delta CPD$ = 1.17 V (at $V_G$ = 0 V) for TABINOL and DPCP, respectively. Both characterizations (IV and KPFM) show an increase of the hysteresis with DPCP exposure, the IV hysteresis is ≈ 2V after TABINOL grafting and ≈ 4V after reaction with DPCP. Note that results presented in figure 11 are not directly comparable with those presented on "naked" Si-NR at figure 10; for TABINOL and DPCP in figure 11 characterizations were done at $V_{DS}$ = -1 V while before chemical modification in figure 10 $V_{DS}$ is -2 V. Nevertheless, the trend is the same as for the sequential method (section 4.1): TABINOL grafting mainly modifies the CPD, while the IV hysteresis remains unchanged, the reaction with DPCP mainly increases hysteresis in both the IV curve and the $V_G$-dependency of CPD.

## 5. DETECTION OF SARIN WITH TABINOL-MODIFIED SI-NR FET

### 5.1. Set up

The detailed procedure for the fabrication of the Si-NR FET used for sarin test are described in[27]. The TABINOL modified Si-NR FETs are connected onto a pluggable plastic card by gold wire bonding to source and drains electrodes and with silver paste deposited onto the back for gate electrode. Three of these sensors were place in a 3-way tight fluidic cell with BNC connectors and connected to three independent detectors where drain-source and gate-source voltages are applied while the drain source current is measured as a function of time (See Supporting Information). The prototype is mainly divided into three parts: a microcontroller, a functionalized SiNW-FET as transductor, and digital-analog converter interfacing the FET with a computer. The detectors are able to measure very low current over a range of six decades. The common range of measurement for our devices is $10^{-10}$ A - $10^{-4}$ A.

### 5.2. Results

A sampling cycle begins with a calibration step which consists in sweeping drain-source current versus gate voltage of the transistor at a constant bias voltage $V_{DS}$ = −1V. Then, the minimum value of the off-current is identified with the corresponding gate voltage ($V_{BG}$). This value of $V_{BG}$ is named $V_0$, and is then applied to monitor $I_{DS}$ as a function of time. It has been shown previously that reaction with OPs causes a shift to more positive $V_{BG}$ with an average variation $\Delta V_m$ of (7.3V ± 3.5V). A series of tests with a real chemical warfare nerve agent, the Sarin, have been carried out with TABINOL functionalized SiNW-FET on the portable detector[25]. Exposure to Sarin vapors induced a modification of the transfer curves as observed with the nerve agent simulants described in this article. The $I_{DS}$–$V_{BG}$ curves are shifted to more positive gate-voltage bias by a few volts (~5V) (Figure 12a). The portable detector sets automatically $V_{BG}$ to -7.5 V as operating voltage. Figure 12b shows a stable $I_{DS}$, at this gate voltage, before exposure to Sarin vapors (40 ppm measured with ppbRAE detector from RAE Systems). After exposure, $I_{DS}$ rise after few seconds with a steep current increase. A plateau is reached within few minutes. The prototype is able to accurately measure $I_{DS}$ down to hundreds of pA. The very stable value of a few pA measured before exposition and the $I_{DS}$ value measured just after the exposure are lower than the limit of detection of the prototype and are inaccurate but this is not detrimental to efficient Sarin detection.

Two factors could explain the longer dead time observed for Sarin compare to DPCP exposure. First, the flow cell used for Sarin exposure was not optimized and is possibly affected by fluidic problems. Secondly, Sarin could exhibit a lower reactivity than DPCP due to the higher bond dissociation energy ($D^0$ at 298 K) of P-F bond (439 kJ.mol$^{-1}$, Sarin) when compared to P-Cl bond (289 kJ.mol$^{-1}$, DPCP). However, the detection is fast with a high on/off ratio, which demonstrates the ability of this technique to detect efficiently traces of Sarin.

## 6. CONCLUSIONS

In summary, we have performed surface characterizations of a sarin detector by KPFM, KP and ToF-SIMS. Simultaneous current-voltage characterization and KPFM mapping on nerve agent sensor based on chemical reaction occurring on SiNR-FET surface at each chemical steps, were achieved. By this approach, we observed clear correlations of the hysteresis between the CPD measured on the Si-NR and the transfer curve of the transistor. More important CPD values were obtained under biased source-drain electrodes and can be explained by the presence of charge carriers into the Si-NR. Comparison between KPFM measurements on the Si-NR FET and KP measurements on large Si surface reveals that they are qualitatively in agreement: a significant increase of the CPD after the TABINOL grafting and a weak decrease after the reaction with DPCP. ToF-SIMS analysis showed the presence of OP molecules essentially localized on the Si-NR after the SiNR-FET exposure, confirming the effectiveness and selectivity of the sensor. Finally a sensitive prototype exposed to real Sarin vapors and based on the Si-NR FET was successfully demonstrated. This work shows that KPFM is a powerful tool in order to understand the electrical behavior of devices based on chemical surface modifications, and paves the way (i) to an analyze in function of the molecular surface coverage due to the sensibility of these surface characterizations and (ii) to the development of commercial sensitive, compact, portable, low-cost, and low-consumption sensors based on the functionalized Si-NR by OP sensors.

ASSOCIATED CONTENT

**Supporting information**. The Supporting Information is available: Fabrication of Nano-ribbon Si FET, grafting of TABINOL, DPCP exposure protocol, KPFM side capacitance corrections, ToF-SIMS: table, KP measurements on samples exposed to air, test of sensors with Sarin.

AUTHOR INFORMATION

**Corresponding Author**

* E-mail: stephane.lenfant@iemn.univ-lille1.fr. Phone : +33 3 20 19 79 07

§ Present address: Centre de Développement des Technologies Avancées, Division Microélectronique et Nanotechnologie, CDTA/DMN, Cité 20 août 1956 Baba Hassen, A-16081, Algérie.

ACKNOWLEDGEMENTS

We thank Dr. Nicolas Nuns (Unité de Catalyse et Chimie du Solide – University of Lille CNRS) for his assistance and advice for the ToF-SIMS analyses. Bruno Simonnet (ID3 Technologies) and Dominique Boichon (NBCSys) are acknowledged for their help in the device fabrication for experiments on real Sarin gas; Dominique Deresmes (IEMN) for his assistance on KPFM. This work has been financially supported by ANR, grant n°ANR-10-CSOSG-003, project CAMIGAZ and the French RENATECH network.

a) 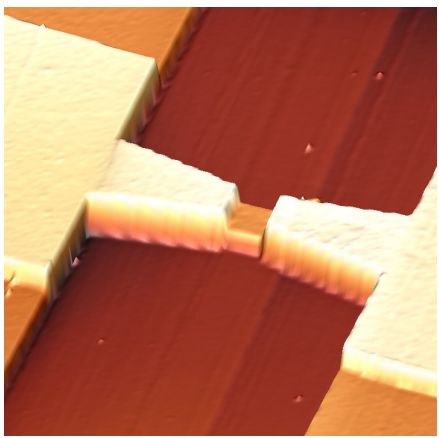

b) 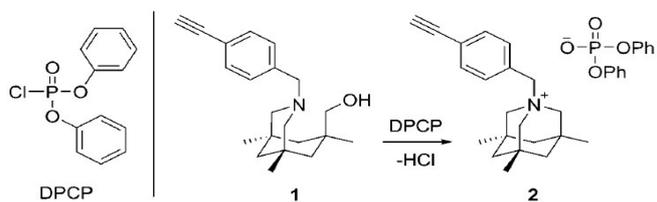

**Figure 1.** (a) TM-AFM image a SiNR-FET (70 nm thick, 4 µm length and 4 µm width) and its source and drain electrodes fabricated from SoI wafers. (b) The Si-NRs were functionalized by covalent grafting with a thermal hydrosilylation of 1 onto HF treated Si-NRs. Compound 1 referred as TABINOL converts into compounds 2 upon exposure to simulant of nerve agent (DPCP).

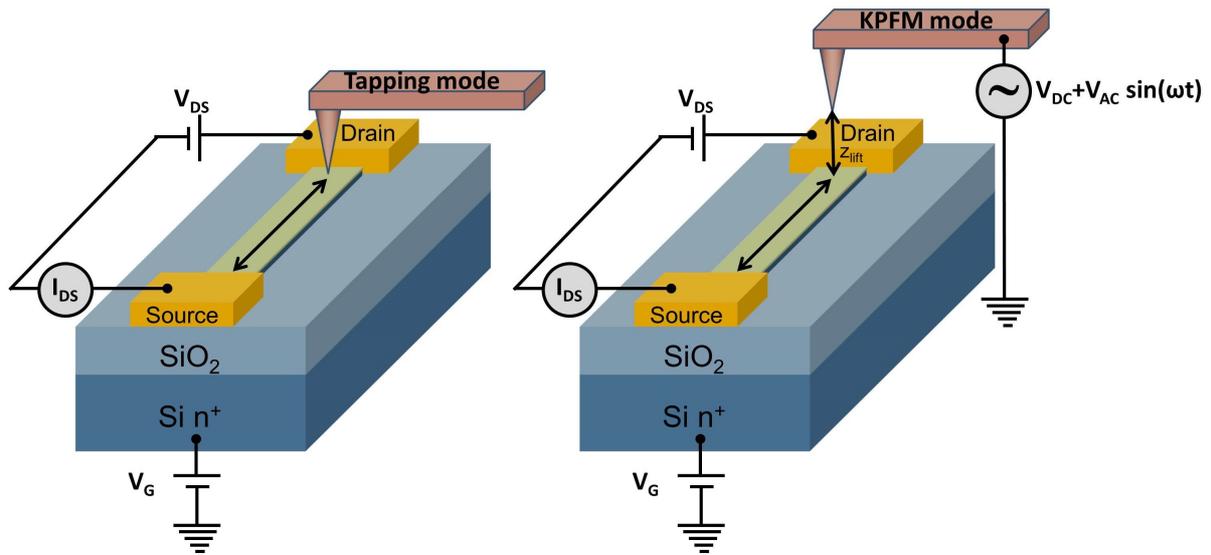

**Figure 2**. Schematics of the measurement setup by a standard two-pass procedure: (a) In the first linear scan, the topography of the line is acquire in tapping mode. (b) In the second linear scan on the same line, KPFM signal is acquire in a lift mode. The Si-NR FET is polarized with a source drain voltage $V_{DS}$ and a back gate voltage ($V_G$). Simultaneously of the KPFM characterization, the current through the Si-NR ($I_{DS}$) is measured.

|                | Water Contact Angle (°) | Thickness (Å)        |
|----------------|-------------------------|----------------------|
| TABINOL        | 66                      | 16.5                 |
| TABINOL + DPCP | 60                      | 17.8 (1h) - 19.3 (2h)|

**Table 1**. Water contact angles and thickness values as measured on large pieces (1 cm$^2$) of treated Si surfaces.

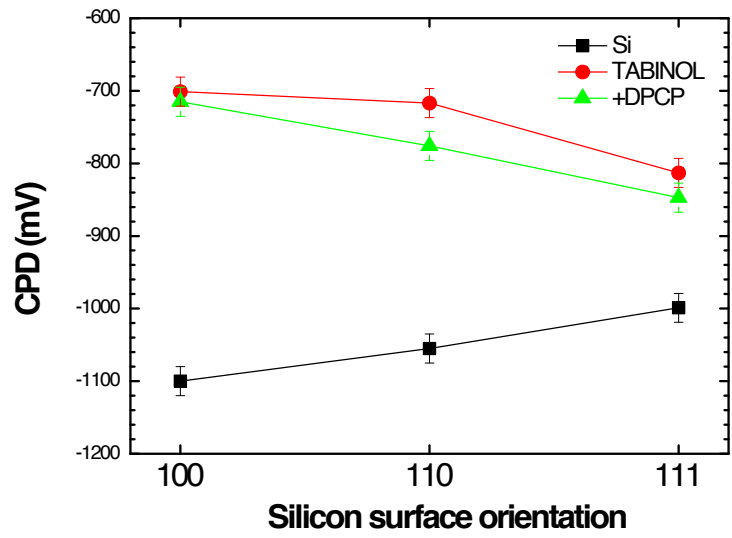

**Figure 3**. CPD values measured for the naked Si surface, and Si surface treated with TABINOL and reacted with DPCP for 3 silicon surface orientations.

|  | CPD by KPFM (mV) | CPD by KP (mV) |
|---|---|---|
| « Naked » Si | +710 | -1100 |
| Si TABINOL | +1060 | -701 |
|  | $\Phi_{TAB} \sim$ **+ 350** | $\Phi_{TAB} \sim$ **+ 399** |

**Table 2**. KP and KPFM characterizations on large surfaces after and before TABINOL grafting on silicon (same silicon than for Si-NR fabrication). Relative variations of the CPD with the TABINOL grafting $\Phi_{TAB}$ are also given.

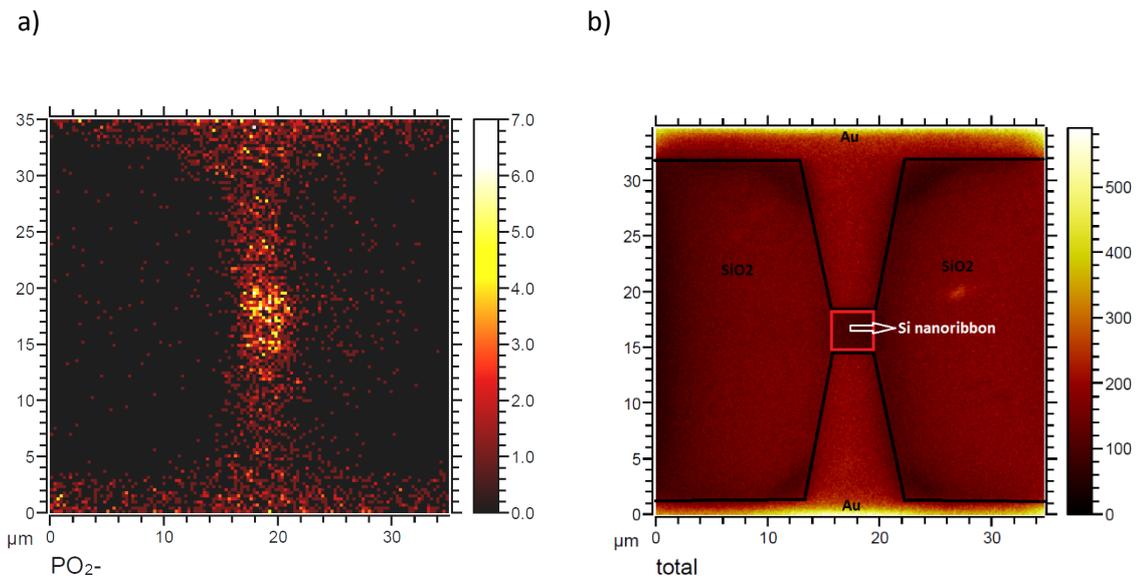

**Figure 4**. (a) High resolution Tof-SIMS images (in negative mode) of $PO_2^-$ ions observed after exposure to DPCP. (b) High resolution Tof-SIMS image (in positive mode) of total sum of fragment ions showing the device geometry.

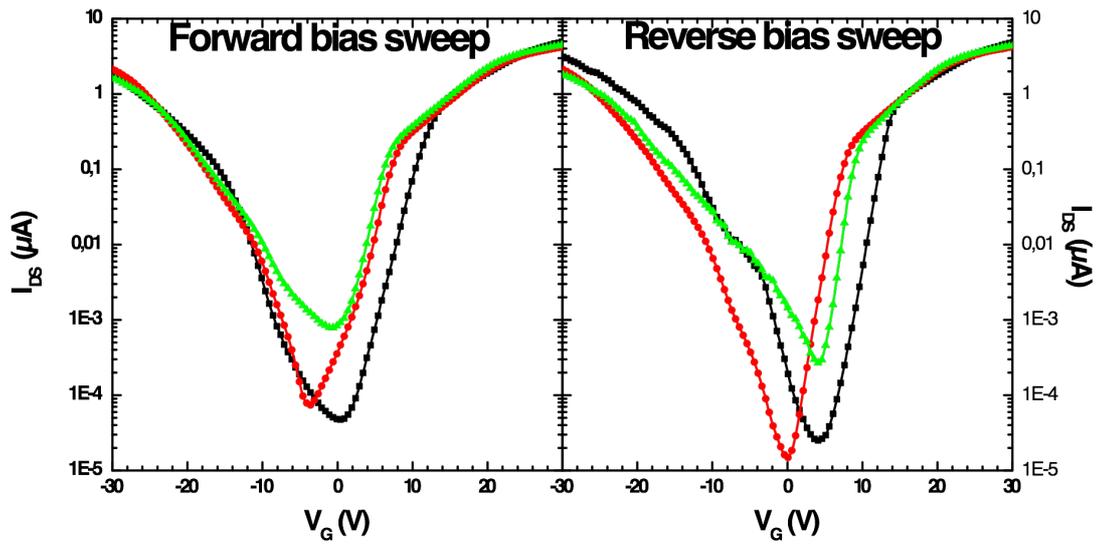

**Figure 5.** Typical $I_{DS}$-$V_G$ curves at $V_{DS}$=-4V measured on 4 x 4 µm² Si-NR FET with "naked" Si (—■—); Si-NR functionalized with TABINOL (—●—); and after DPCP exposure (—▲—). Forward bias sweeps from -30V to +30V, reverse bias sweeps from +30V to -30V.

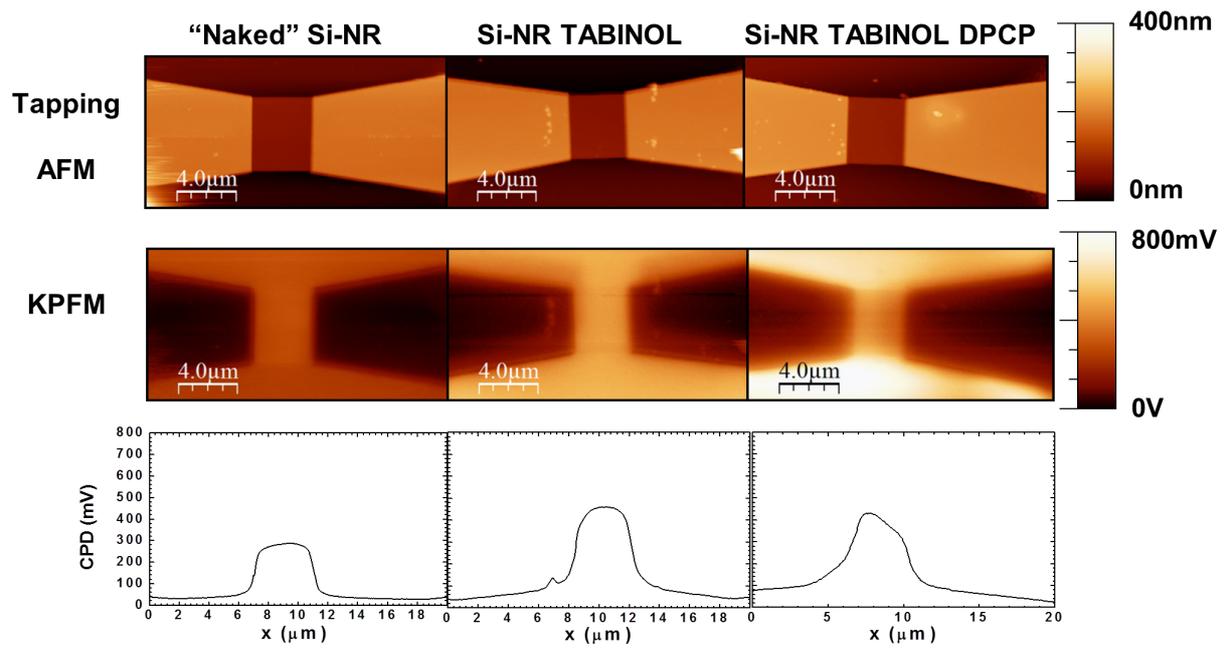

**Figure 6.** Tapping AFM image, KPFM image and experimental CPD profile of Si-NR FET of 4 x 4 µm² (average of the 200 lines on the Si-NR), before and after TABINOL functionalization and after DPCP exposure.

a)

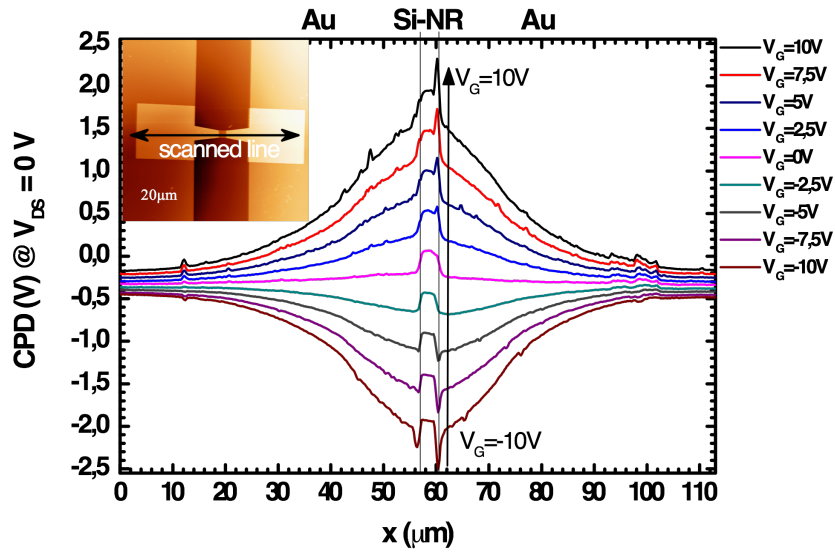

b)

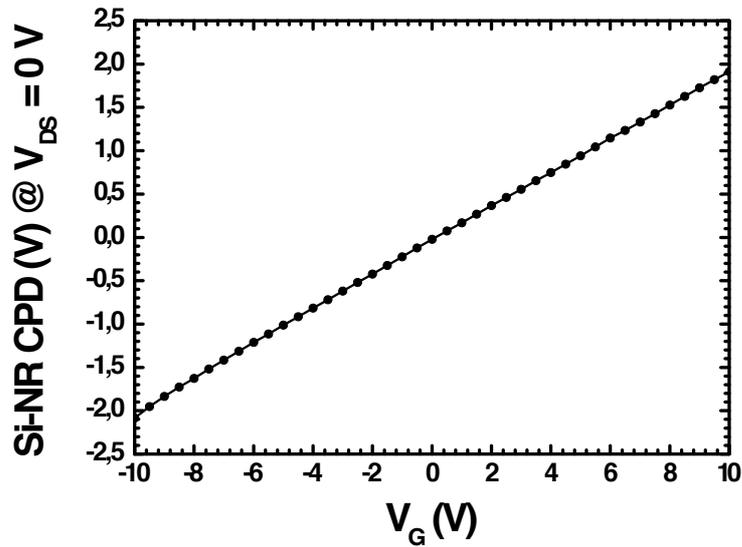

**Figure 7.** (a) Measured CPD profiles measured along the Si-NR FET of 4 x 4 µm² under different applied back gate potentials for a source-drain potential $V_{DS}$ = 0 V. In inset the AFM image of the Si-NR FET studied. The arrow represents the segment Au electrode/Si-NR/Au electrode analyzed by KPFM and presented in the CPD profiles; (b) Measured CPD taken at the middle of the Si-NR between source and drain (from Fig. 7-a).

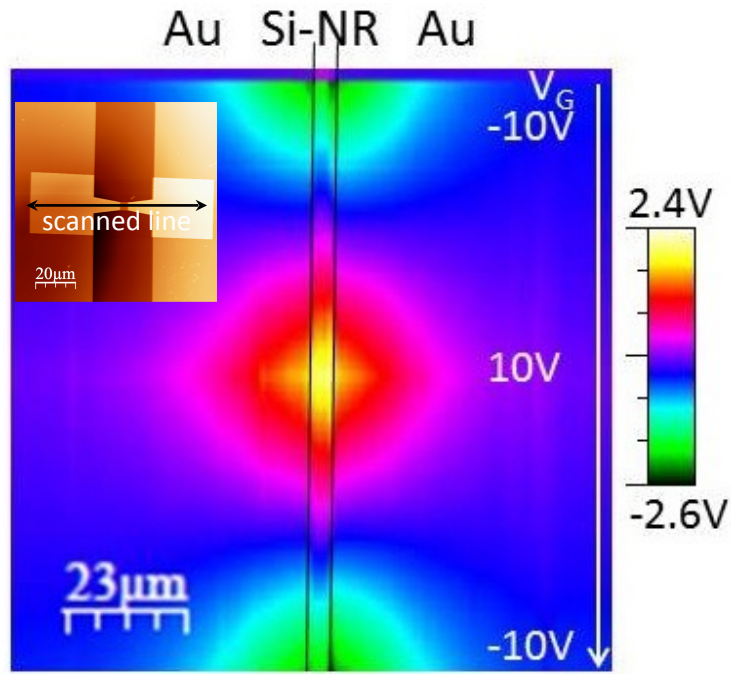

**Figure 8.** KPFM image of the same segment Au electrode/Si-NR/Au electrode presented in inset on the AFM image of the Si-NR FET (function "slow scan axis" disabled in Nanoscope software) in function of the back gate potential ($V_G$). $V_G$ is varying from -10V (line on top of the image), +10V (line in the middle of the image), to -10V (last line of the image) at $V_{DS}$ = 0 V.

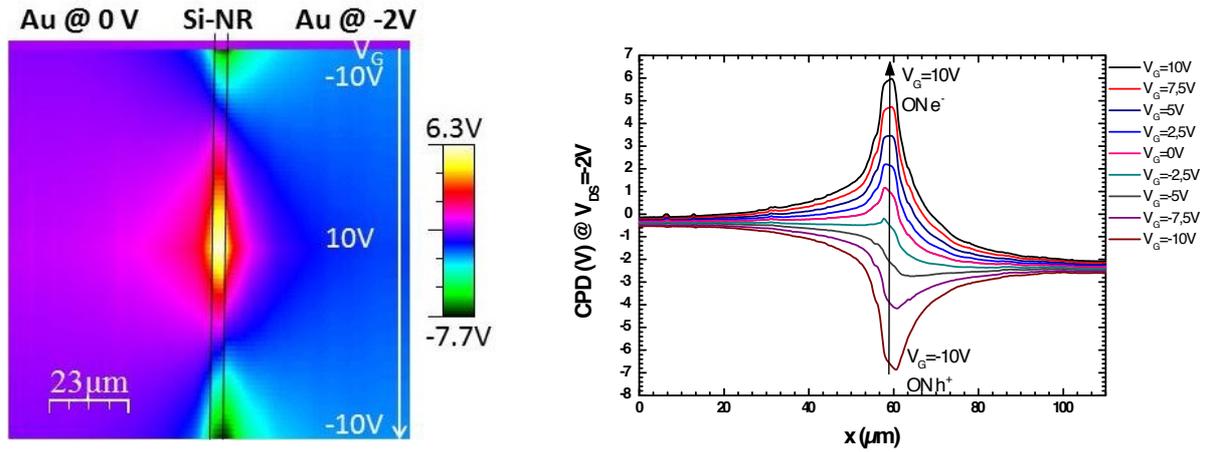

**Figure 9.** Uncorrected CPD profiles measured by KPFM of the segment Au electrode/Si-NR/Au electrode presented in inset in figure 8 for $V_{DS}$ = -2 V. (a) KPFM image of the same segment Au electrode/Si-NR/Au electrode presented in inset at the figure 8 (function "slow san axis" disabled in Nanoscope software) in function of the back gate potential ($V_G$). $V_G$ is varying from -10V (line on top of the image), +10V (line in the middle of the image), to -10V (last line of the image). (b) CPD profiles measured along the Si-NR FET of 4 x 4 µm² under different applied back gate potentials for a source-drain potential $V_{DS}$ = -2 V.

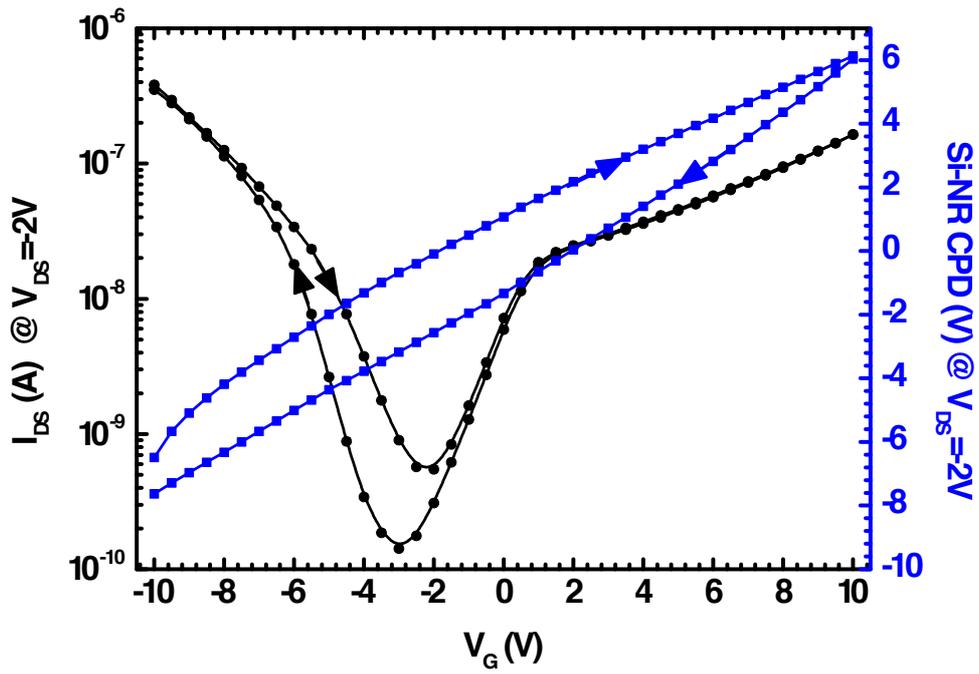

**Figure 10.** $I_{DS}$-$V_G$ curves at $V_{DS}$ = -2 V measured on the Si-NR FET 4 x 4 µm² before chemical functionalization ("naked" Si-NR) (■); compared with the CPD dependence with $V_G$ measured simultaneously on the Si-NR surface in function of the back gate potential ($V_G$) (●).

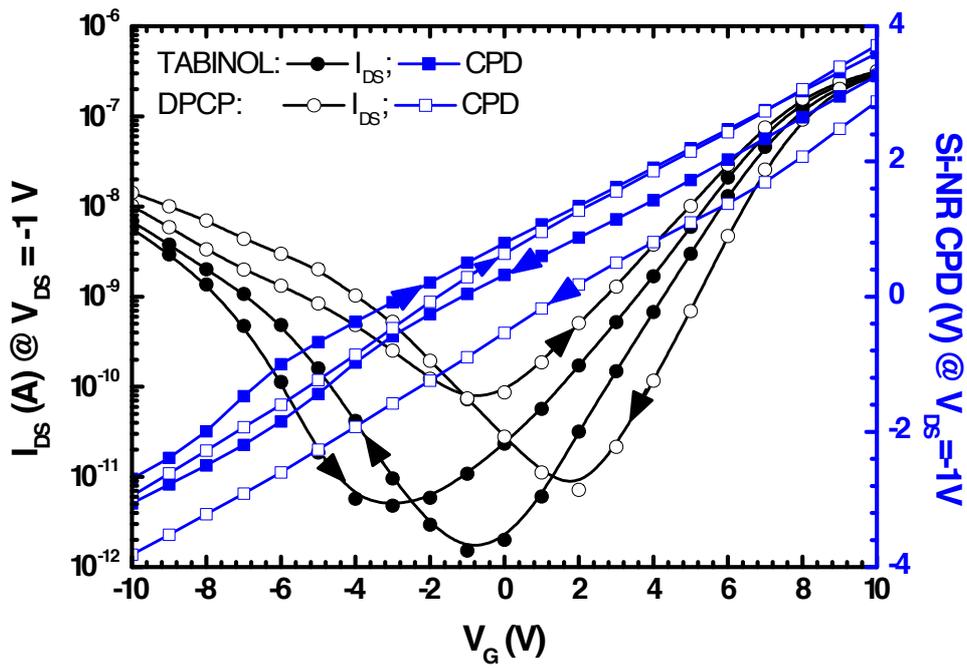

**Figure 11**. $I_{DS}$-$V_G$ curves at $V_{DS}$ = -1 V measured on the Si-NR FET (4 x 4 µm²) with TABINOL functionalization before (solid line) and after (dots line) chemical exposition to DPCP. These curves are compared with the CPD dependence with $V_G$ measured simultaneously on the Si-NR surface in function of the back gate potential ($V_G$).

a) 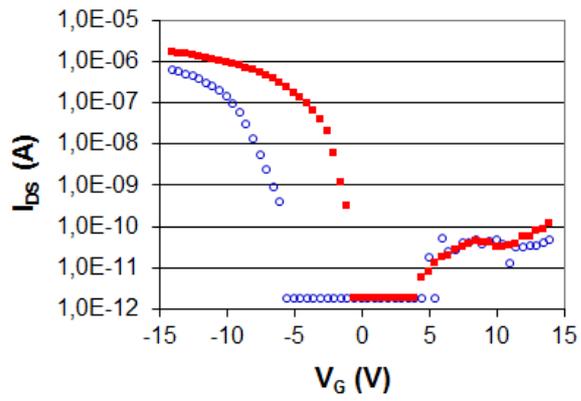 b) 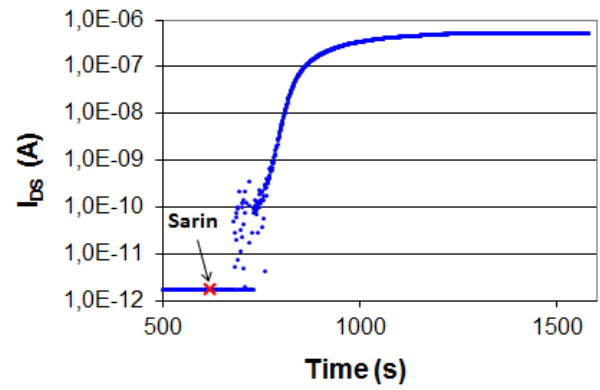

**Figure 12.** (a) Transfer curves of TABINOL modified SiNR-FET before (blue circle) and after (red square) exposure to Sarin vapors; (b) $I_{DS}$ as a function of time ($V_G$=-7.5V, $V_{DS}$=-0.5V) Sarin vapors introduced at *t*=620s.

**TOC GRAPHIC**

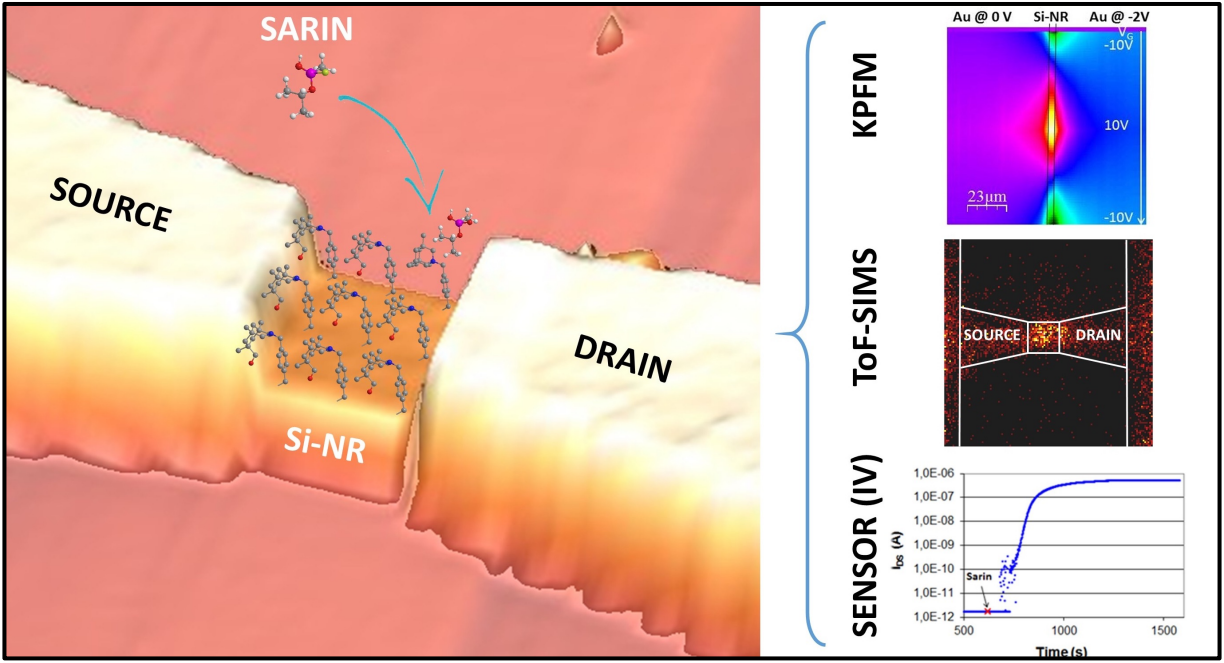

# Physical Study by Surface Characterizations of Sarin Sensors on the Basis of Chemically Functionalized Silicon Nanoribbon Field Effect Transistor

K. Smaali[1], D. Guérin[1], V. Passi[1], L. Ordronneau[2], A. Carella[2], T. Mélin[1], E. Dubois[1], D. Vuillaume[1], J.P. Simonato[2] and S. Lenfant[1,*]

[1] IEMN, CNRS, Avenue Poincaré, Villeneuve d'Ascq, F-59652cedex, France.

[2.] CEA, LITEN/DTNM/SEN/LSIN, Univ. Grenoble Alpes, MINATEC Campus, F-38054 Grenoble, France.

* Corresponding author: stephane.lenfant@iemn.univ-lille1.fr

## SUPPORTING INFORMATION

## 1. NANO-RIBBON SI FET

Silicon nanoribbons were fabricated using e-beam lithography and reactive-ion etching (RIE) on SoI (Silicon on Insulator) substrates. The top silicon (p-type) layer, 70 nm thick, was doped with $10^{15}$ cm$^{-3}$ boron impurities. The thickness of the silicon dioxide is 140 nm on a silicon substrate (10 Ω.cm) serving as the bottom gate. A thin layer of 45 nm of negative resist (HSQ FOX 12) was deposited by spin-coating on clean and deoxidized SoI substrate. The resist was developed by dipping the substrate in tetramethyl ammonium hydroxide (TMAH) 25 % for 1 min, after e-beam exposition. Combining HSQ resist as mask with RIE with a SF$_6$ 15 sccm / N$_2$ 10 sccm / O$_2$ 10 sccm plasma (10 mT, 50 W, 80 s in Plasmalab system from Oxford Instrument), we obtained nanoribbons with different lengths and widths (4x4 µm, 4x1 µm, 2x1µm, 2x0.2µm) connected to 30 µm x 30 µm silicon squares for the source and drain contacts. Finally, HSQ resist was removed by wet etching (HF 1%, 1 min). This step reduced the thickness of exposed silicon dioxide to around 134 nm as measured by ellipsometry. The large (100 µm x 100 µm) metal contacts on the 30 µm x 30 µm silicon pads were patterned by e-beam lithography with a double layer resist (copolymer EL 10 % - MMA 17.5 % and PMMA 3 % 495 K), 610 nm and 60 nm thick, respectively, deposited by spin coating. After e-beam exposition, the resist was developed with a solution of 1/3 Methyl isobutyl ketone (MIBK) and 2/3 isopropanol for 60 s. Metal layer (titanium 10 nm and gold 100 nm) was deposited using an e-beam evaporator in high vacuum. The lift-off was carried out in acetone bath.

## 2. GRAFTING OF TABINOL

The same procedure was followed for the functionalization of the SoI chip bearing silicon nanoribbons of various dimensions and for the silicon wafer used for contact angle measurements, ellipsometry and KP measurements. Deoxidation of subtrates and hydrosilylation reactions were realized in a nitrogen glovebox ($H_2O$ and $O_2$ < 1 ppm). Schlenk glassware was dried in an oven at 120°C for 2 days. Hydrofluoric acid (1 % HF in water), mesitylene and deionized water were degassed for 15 min by nitrogen bubbling. Mesitylene was stored over 4 Å molecular sieves for 48h before use. The substrates were sonicated 5 min in acetone then 5 min in isopropanol. They were immersed in the degassed 1 % HF solution for 20 seconds and then thoroughly rinsed with degassed deionized water. After drying under nitrogen flow, they were introduced in a schlenk flask containing a $10^{-3}$ M solution of TABINOL in anhydrous mesitylene (3 mg in 10 mL). The schlenk was sealed then transferred in an oil bath at 160 °C for 2 h. The chip and the Si wafer were sonicated 2 min successively in toluene, acetone then isopropanol, and finally they were dried under nitrogen flow.

## 3. DPCP EXPOSURE PROTOCOL

The functionalized silicon nanoribbons were fully characterized before and after 1 h exposure to DPCP vapors. In a typical experiment, two drops of DPCP were deposited in a 50 mL closed Petri dish to generate a stabilized vapor pressure of DPCP, the chip was then introduced carefully so that it was not in direct contact with the liquid organophosphorus compound but only exposed to the vapor for one hour. The chip was again fully characterized electrically after being removed from the DPCP vapors.

## 4. KPFM: SIDE CAPACITANCE CORRECTIONS

We describe here the treatment of side capacitance effects taking place in KPFM experiments. From [H. O. Jacobs, P. Leuchtmann, O. J. Homan, and A. Stemmer, J. Appl. Phys. 84, 1168 (1998)], the CPD measured by KPFM is an averaged value of the potentials $V_i$ probed within the tip-cantilever/substrate capacitance, which can be expressed in the case of amplitude-modulation KPFM as : CPD=$\sum \frac{dC_i}{dz} V_i / (\sum \frac{dC_i}{dz})$. We consider the potentials $V_i$ below the tip apex (i=1), tip cone (i=2, i=3) and cantilever (i=4) according to Figure SI-1a, yielding the following expression for the KPFM signal: CPD = $\alpha_1 V_1 + \alpha_2 V_2 + \alpha_3 V_3 + \alpha_4 V_4$, in which the $\alpha_i$ coefficients are determined experimentally, as done in Ref. [D. Brunel, D. Deresmes, and T. Mélin, Appl. Phys. Lett. 94 223508 (2009)], and $\alpha_1+\alpha_2+\alpha_3+\alpha_4=1$. A top view schematics of the cantilever is represented at the scale of the NR-FET structure in Figure SI-1b. It illustrates that the NR-FET device is imaged in KPFM at a sub-100nm (local) scale by the tip apex capacitance gradient $C'_1$, and, in parallel, by the tip cone capacitance gradient components $C'_2$ and $C'_3$ (at a few µm – 10 µm scale) and by the cantilever (with integration at a few 10 µm scale) via $C'_4$.

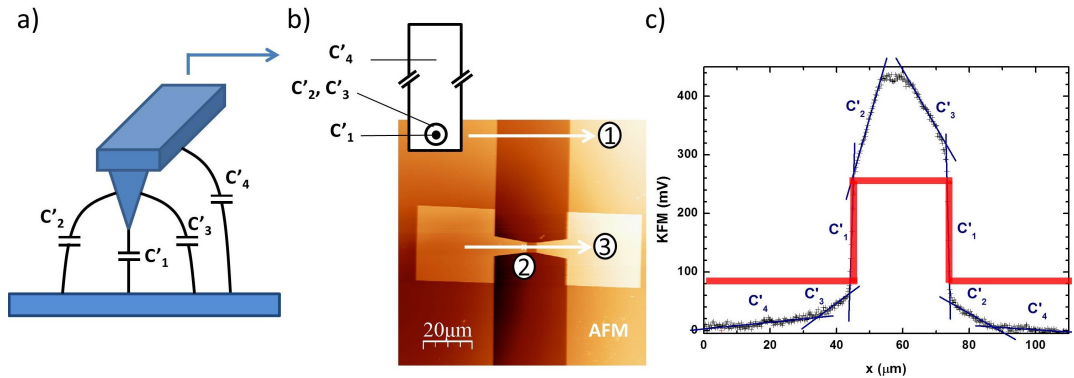

**Figure SI-1.** (a) Schematics representation of side capacitances between tip and surface (see text for full description); (b) Two different displacements studied (noted ① and ②) of the tip on the surface; (c) Ideal CPD signal (red line) compared to the experimental CPD signal for the displacement ②.

The determination of $\alpha_1$, $\alpha_2$, $\alpha_3$ and $\alpha_4$ is carried out as follows. $\alpha_4$ is obtained by placing the tip at the position shown in Figure SI-1b, with $C'_1$, $C'_2$ and $C'_3$ over the Au metallization, while $C'_4$ is left above the substrate SiO$_2$ layer. At this position, the surface potential $V_4$ felt by $C'_4$ thus follows the FET backgate bias $V_G$. $\alpha_4$ is therefore obtained as the proportionality coefficient between the recorded CPD and $V_G$, yielding $\alpha_4 \sim 12 \pm 1$ %. $\alpha_1$, $\alpha_2$, and $\alpha_3$ are then determined by scanning the tip across the device Au metallization pattern (arrow labelled 1 in Figure SI-1b) at $V_G = 0$. Ideally, the CPD should be a square (two-level) signal associated with the SiO$_2$ layer opening (with width 30 µm) between Au metallization pads. The ideal CPD signal and a plot of the experimental KPFM profile are shown in Figure SI-1c. This enables to identify the weights of the capacitance gradients $C'_1$, $C'_2$ and $C'_3$, when the tip is scanned across the border between the Au metallization pad and the SiO$_2$ layer. $\alpha_1$ is *e.g.* identified from the sharp jump in CPD across this border (with sub 100nm resolution), while the amplitude of $\alpha_2$ and $\alpha_3$ are obtained from the amplitude of the linear shifts of the CPD observed at a few µm scale around the Au/SiO$_2$ border. This enables to calculate the relative weights between C'$_1$, C'$_2$ and C'$_3$, and finally to obtain the full set of $\alpha_i$ values : $\alpha_1$=40±3 % ; $\alpha_2$=27±3 % ; and $\alpha_3$=21±3 %. The difference between $\alpha_2$ and $\alpha_3$ are likely related to the cantilever axis misalignment with the device pattern, as can be observed from the AFM image in Figure SI-1b.

We now consider the situation where the tip is located at the center of the FET NR (position 2 in Figure SI-1b), where we want to extract quantitative information from the KPFM measurements recorded after exposure to TABINOL and DPCP in the main part of the paper. In this configuration, obviously, the C'$_1$ capacitance (tip apex) is only sensitive to the Si NR CPD. Since C'$_2$ and C'$_3$ have an integration radius up to the 10 µm scale (as seen from Figure SI-1c), these capacitances thus "feel" the Si NR CPD, as well the adjacent Au metallization CPD, and the SiO$_2$ CPD. Finally, since the cantilever width (30 µm) is similar to the gap between the Au metallization pads, C'$_4$ is mostly sensitive to the SiO$_2$ layer CPD, and to a lower extent, to the Au metallization CPD. Therefore, the KPFM CPD signal measured at the center of the NR-FET device will be a weighted average of (i) the NR CPD $V_{NR}$; (ii) the Au metallization CPD $V_{Au}$ (which may also depend on the TABINOL and DPCP exposure) ; and (iii) the SiO$_2$ CPD $V_{ox}$ (which linearly varies as a function of the back-gate bias $V_G$).

We start with the decomposition of C'$_2$ and C'$_3$. To do this, we take advantage of the Schottky character of the Au contacts on the 4x4 µm² NR-FET device of Figure SI-1. In this situation, a positive $V_{ds}$ polarization (i.e. $V_{ds}$ applied to the NR-FET right contact, the left contact being grounded) biases the NR at $V_{ds}$ (voltage drop at the left Au/Si contact), while a negative bias leaves the NR at ground (voltage drop at the right Au/Si contact). Using the experimental KPFM data plotted as a function of $V_{ds}$ (data not shown here), we obtain a set of two equations which enable to derive the fraction of $C'_2$ and $C'_3$ probing the NR ($\alpha_{NR}$) and the adjacent Au metallization ($\alpha_{Au}$), respectively. We find $\alpha_{NR} \approx 35\%$, $\alpha_{Au} \approx 30\%$. The fraction of $C'_2$ and $C'_3$ probing the SiO$_2$ CPD ($\alpha_{ox}$) is $\alpha_{ox}=1-\alpha_{NR}-\alpha_{Au} \approx 35\%$. A similar work also enables the determination of the fraction of $C'_4$ probing the SiO$_2$ layer ($\alpha'_{ox} \approx 87\%$) and the Au metallization ($\alpha'_{Au} \approx 13\%$).

From these estimations, we obtain the relationship between the measured CPD signal in the center of the NR of the FET device, as a function of the CPD of the NR, oxide (i.e. back-gate), and Au metallization pads:

$$CPD = \alpha_1 \cdot V_{NR} + (\alpha_2+\alpha_3) \cdot (\alpha_{NR} \cdot V_{NR} + \alpha_{ox} \cdot V_{ox} + \alpha_{Au} \cdot V_{AU}) + \alpha_4 \cdot (\alpha'_{ox} \cdot V_{ox} + \alpha'_{Au} \cdot V_{Au})$$

Numerically, one obtains for the AFM tip at the center of the Si-NR:

$$CPD = 0.57\, V_{NR} + 0.27\, V_{ox} + 0.16\, V_{Au}$$

When used at fixed $V_{Au}$ (i.e. at fixed $V_{ds}$ conditions, and without comparing different chemical steps for the NR-FET device), the variations of the CPD can be thus corrected from the back-gate bias $V_G$ (assuming $V_{ox}=V_G$), and from side capacitance effects, using an effective CPD value:

$$CPD^* = V_{NR} = (CPD - 0.27 V_G - 0.16\, V_{Au})/0.57 = 1.75\, (CPD - 0.27\, V_G - 0.16\, V_{Au})$$

## 5. TOF-SIMS: TABLE

| *Tabinol monolayer* | | *After reaction with DPCP* | |
|---|---|---|---|
| *Mass (m/z)* | *Assignment* | *Mass (m/z)* | *Assignment* |
| 145.05 | $C_9H_9Si^+$ | 145.05 | $C_9H_9Si^+$ |
| 196.17 | $C_{12}H_{22}NO^+$ | 196.17 | $C_{12}H_{22}NO^+$ |
| 294.22 | $C_{21}H_{28}N^+$ | 294.22 | $C_{21}H_{28}N^+$ |
| 312.23 | $C_{21}H_{30}NO^+$ | 312.23 | $C_{21}H_{30}NO^+$ |
| 314.22 | $C_{21}H_{32}NO^+$ | 314.22 | $C_{21}H_{32}NO^+$ |
| | | 62.96 | $PO_2^-$ |
| | | 78.96 | $PO_3^-$ |
| | | 154.98 | $C_6H_4PO_3^-$ |
| | | 170.97 | $C_6H_4PO_4^-$ |
| | | 249.02 | $C_{12}H_{10}PO_4^-$ |

**Table SI-1.** Characteristic fragment ions observed by ToF-SIMS. Mass resolution was >7000 for all selected fragments (>8500 for m/z = 196, 249, 294, 312 and 314)

## 6. KP MEASUREMENTS ON SAMPLES EXPOSED TO AIR.

Experimentally, the Kelvin probe technique measures the contact potential difference ($CPD_{meas}$) between two surfaces brought in close proximity as:

$$CPD_{meas} = (W_s - W_{tip})/e$$

with $W_s$ and $W_{tip}$ the work function of the sample and the tip, respectively, and e the electron charge. But in order to compare directly CPD values with the same definition than KPFM (see Eq. 3 in article), the CPD value presented in KP (CPD) is the opposite of $CPD_{meas}$:

$$CPD = -CPD_{meas}$$

Stability of TABINOL monolayer was verified by placing the functionalized large Si pieces in ambient air for 30 min over duration of day. It was observed that degradation of TABINOL occurred on this exposure to air. The measured CPD by KP shows an increase by about 50 mV after a day. It is likely that due to the steric hindrance of the TABINOL not all the surface hydride Si-H sites are bonded with the TABINOL. These hydrid bonds are very sensitive to the ambient air and they can form hydroxyl (OH) groups when kept in ambient air. Thus when the TABINOL grafted samples are kept in ambient air the hybrid groups form hydroxyl groups and slowly lead to the reformation of a native oxide at the TABINOL/Si interface. Similarly, bare silicon surfaces were measured by KP at various stages, i.e., immediately after removal of the native oxide by HF-5% treatment, and when exposed to air for intervals of 6, 12, 18, 24 and 48h, respectively. We observed a gradual reduction of the CPD values indicating the growth of an oxide layer. For example, the CPD decreases by about 200 mV after 12h of exposure and about 300 mV after 48h. The values of the CPD stabilize after 48h indicating the presence of a stable native oxide layer. The values of CPD did not show much variation after exposed to air for 48h. Thus, these features may partly explain the difference of $\Phi_{TAB}=CPD_{TAB} - CPD_{ref}$ when measured by KP (in vacuum) - about 400 mV, and by KPFM in a $N_2$ purged ambient (likely with a certain amount of residual oxygen).

## 7. TEST OF SENSORS WITH SARIN

Tedlar bags are commonly used gas sampling bags. Tedlar bags of 20 L were filled in advance with Sarin, determined by weight. The concentration of Sarin was expressed in ppm and was calculated from the weight of CWA and the air volume in the bag and measured with ppbRAE detector.

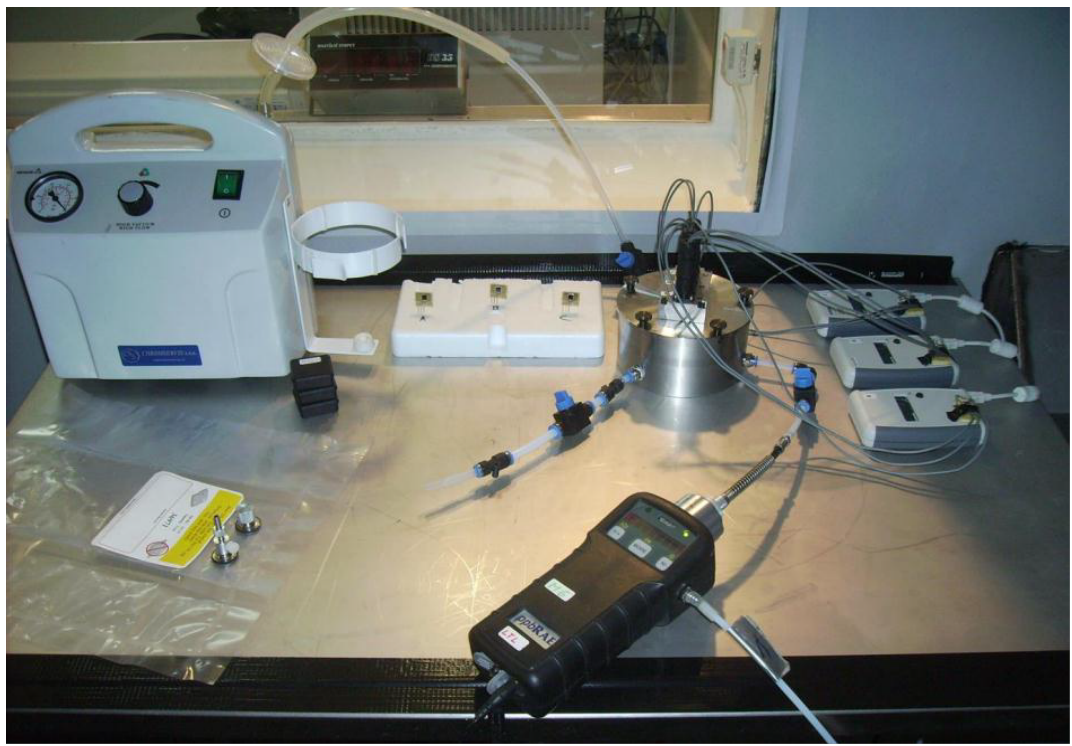

**Figure SI-2.** Experimental setup picture

The basic protocol consisted of the following steps:

1. Insert three sensors on the underside of the lid of the flow cell.

2. Close the lid and fasten the five screws.

3. Connect all components to the flow cell.

4. Open valve to vacuum pump to create vacuum, down to -0.8 bar.

5. Close valve to vacuum pump.

6. Open valve to Tedlar bag with sarin. Sarin flows into the flow cell.

7. Switch on flow to the ppbRAE detector.

8. Keep flow for at least 10 min.

9. Close valve to Tedlar bag.

10. Allow fresh air from the room to enter the flow cell. Keep the flow with the ppbRAE detector on.

11. Flush system with air until the ppbRAE detector shows background value.

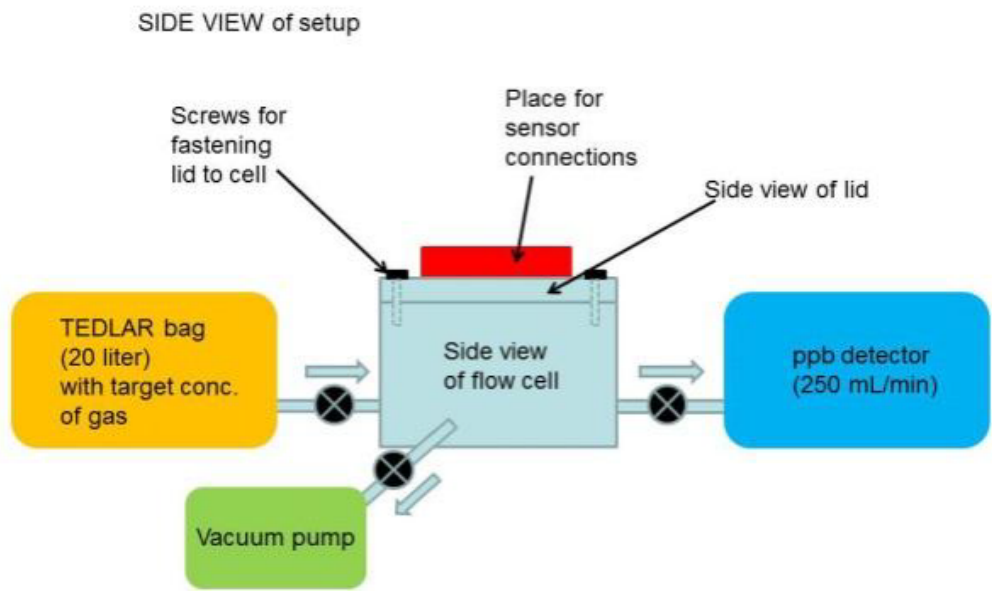

**Figure SI-3.** Experimental setup description